# Efficient semiquantum key distribution based on single photons in both polarization and spatial-mode degrees of freedom

Tian-Yu Ye*, Mao-Jie Geng, Tian-Jie Xu, Ying Chen

College of Information & Electronic Engineering, Zhejiang Gongshang University, Hangzhou 310018, P.R.China

**Abstract:** In this paper, we propose an efficient semiquantum key distribution (SQKD) protocol which is based on single photons in both polarization and spatial-mode degrees of freedom. This protocol is feasible for a quantum communicant distributing a random private key to a classical communicant. This protocol needn't require the classical communicant to use any quantum memory or unitary operation equipment. We validate the complete robustness of the transmissions of single photons between two communicants. It turns out that during these transmissions, if Eve wants not to be detected by two communicants, she will obtain nothing useful about the final shared key bits. Compared with Boyer *et al.*'s famous pioneering SQKD protocol (Phys Rev Lett, 2007, 99:140501), this protocol has double quantum communication capacity, as one single photon with two degrees of freedom for generating the key bits can carry two private bits; and this protocol has higher quantum communication efficiency, as it consumes less qubits for establishing a private key of the same length. Compared with the only existing SQKD protocol with single photons in two degrees of freedom (Int J Theor Phys, 2020, 59: 2807), this protocol has higher quantum communication efficiency.

## 1 Introduction

Quantum cryptography, invented by Bennett and Brassard [1] when they put forward the first quantum key distribution (QKD) scheme in the year of 1984, is famous for its theoretically unconditional security. It is well known that QKD aims to establish a random private key between two remote communicants through the law of quantum mechanics. In the year of 2007, Boyer *et al.* [2-3] invented a novel branch for quantum cryptography named as semiquantum cryptography, which permits the classical communicants to have limited quantum capabilities. Obviously, semiquantum cryptography allows the classical communicant not to be involved into the preparation and measurement of quantum superposition states and quantum entangled states. Consequently, it is beneficial for the classical communicant to reduce the burdens of quantum state preparation and measurement. Soon after the birth of semiquantum cryptography, many researchers quickly threw their enthusiasms onto the study of semiquantum key distribution (SQKD). As a result, numerous SQKD schemes [4-13] have been constructed, such as the ones based on single photons [4-7], Bell entangled states [8-10], three-qubit entangled states [11,12], four-particle cluster states [13], and so on.

In the quantum cryptography protocols based on single photons [14-17], the quantum communication capacity usually increases along with the number of degrees of freedom for single photons. In order to enlarge the quantum communication capacity for SQKD, in the year of 2020, we put forward a novel SQKD protocol with single photons in both polarization and spatial-mode degrees of freedom [18]. It is popularly accepted that quantum communication efficiency is a great concern for a quantum cryptography protocol. In this paper, for improving the quantum communication efficiency of the SQKD protocol in Ref.[18], we propose an efficient SQKD protocol with single photons in the same degrees of freedom by increasing the number of kinds of initial quantum states.

## 2 Preliminary knowledge

It is popularly known that two nonorthogonal measuring bases in the polarization degree of freedom can be represented as $Z_P = \{|H\rangle, |V\rangle\}$ and $X_P = \{|R\rangle, |A\rangle\}$, where

$$|R\rangle = \frac{1}{\sqrt{2}}(|H\rangle + |V\rangle), \quad |A\rangle = \frac{1}{\sqrt{2}}(|H\rangle - |V\rangle). \quad (1)$$

Here, $|H\rangle$ and $|V\rangle$ are the horizontal and the vertical polarizations of photons, respectively. Likewise, two nonorthogonal measuring bases in the spatial-mode degree of freedom can be described as $Z_S = \{|b_1\rangle, |b_2\rangle\}$ and $X_S = \{|s\rangle, |a\rangle\}$, where $|b_1\rangle$ and $|b_2\rangle$ are the upper and the lower spatial modes of photons, respectively; and

$$|s\rangle = \frac{1}{\sqrt{2}}(|b_1\rangle + |b_2\rangle), \quad |a\rangle = \frac{1}{\sqrt{2}}(|b_1\rangle - |b_2\rangle). \quad (2)$$

Then, we can use [14]

$$|\phi\rangle = |\phi\rangle_P \otimes |\phi\rangle_S \quad (3)$$

to depict a single-photon state in both polarization and spatial-mode degrees of freedom. Here, $|\phi\rangle_P \in \{|H\rangle, |V\rangle, |R\rangle, |A\rangle\}$ is the

single-photon state in the polarization degree of freedom, while $|\phi\rangle_S \in \{|b_1\rangle, |b_2\rangle, |s\rangle, |a\rangle\}$ is the single-photon state in the spatial-mode degree of freedom.

## 3  The designed SQKD protocol

Suppose that quantum Alice wants to distribute a random private key to classical Bob via the quantum channel. The following SQKD protocol is designed to make it possible. Here, the CTRL operation refers to sending back the received single photon directly; and the SIFT operation refers to measuring the received single photon with the $Z_P \otimes Z_S$ basis, recording the measurement result and resending a fresh one in the same state as found.

**Step 1:** Alice generates $1.5n(1+\delta)$ single photons in both polarization and spatial-mode degrees of freedom randomly in the $Z_P \otimes Z_S$ basis. Then, Alice produces $0.5n(1+\delta)$ ones randomly in the $Z_P \otimes X_S$ basis, $0.5n(1+\delta)$ ones randomly in the $X_P \otimes Z_S$ basis and $0.5n(1+\delta)$ ones randomly in the $X_P \otimes X_S$ basis, respectively. Afterward, Alice randomly reorders all single photons in her hand. Finally, Alice sends them to Bob one by one. Note that after Alice sends the first one to Bob, she sends another one only after receiving the previous one. Here, $\delta > 0$ is some fixed parameter.

**Step 2:** For each coming single photon, Bob randomly chooses to SIFT or CTRL. Note that there are $0.75n(1+\delta)$ single photons Alice prepared in the $Z_P \otimes Z_S$ basis and Bob chose to SIFT.

**Step 3:** Bob publishes which single photons he chose to SIFT. Alice publishes which single photons were prepared in the $Z_P \otimes Z_S$ basis. Alice uses her corresponding preparing basis to measure the single photons Bob chose to CTRL and the $Z_P \otimes Z_S$ basis to measure the single photons Bob chose to SIFT. For security check, Alice randomly chooses $0.25n(1+\delta)$ single photons among the ones she prepared in the $Z_P \otimes Z_S$ basis and Bob chose to SIFT, and tells Bob the positions of these chosen ones. For simplicity, these chosen ones are called as the $Z_P \otimes Z_S$ _SIFT_CHECK single photons.

For the single photons Bob chose to CTRL, Alice computes the error rate through comparing their initial prepared states with her own measurement results on them. For the single photons Bob chose to SIFT and Alice prepared in the $Z_P \otimes X_S$ basis, the $X_P \otimes Z_S$ basis or the $X_P \otimes X_S$ basis, Alice requires Bob to tell her his measurement results and computes the error rate through comparing Bob's measurement results on them with her own measurement results and their initial prepared states. For the $Z_P \otimes Z_S$ _SIFT_CHECK single photons, Alice also asks Bob to tell her his measurement results and also calculates the error rate by comparing Bob's measurement results on them with her own measurement results and their initial prepared states. If all of the above error rates are low enough, the communication will be continued; otherwise, the communication will be halted.

**Step 4:** Alice and Bob select the first $0.5n$ single photons from the remaining $0.5n(1+\delta)$ ones Alice prepared in the $Z_P \otimes Z_S$ basis and Bob chose to SIFT to generate the final shared key bits according to the following rule: if the state of the $t$ th single photon is $|H\rangle \otimes |b_1\rangle$, then $(k_{2t-1}, k_{2t}) = (0,0)$; if the state of the $t$ th single photon is $|H\rangle \otimes |b_2\rangle$, then $(k_{2t-1}, k_{2t}) = (0,1)$; if the state of the $t$ th single photon is $|V\rangle \otimes |b_1\rangle$, then $(k_{2t-1}, k_{2t}) = (1,0)$; and if the state of the $t$ th single photon is $|V\rangle \otimes |b_2\rangle$, then $(k_{2t-1}, k_{2t}) = (1,1)$. Here, $k_{2t-1}$ and $k_{2t}$ are the $2t-1$ th and the $2t$ th bits of the final shared key, respectively, and $t = 1, 2, \cdots, 0.5n$.

It concludes the description of the proposed SQKD protocol. It is worthy of emphasizing that the classical communicant, Bob, is not required to use any quantum memory or unitary operation equipment. In addition, some important differences between this protocol and the SQKD protocol of Ref.[18] are worthy of being pointed out: (1) in the former, Alice generates single photons in two degrees of freedom randomly in the $Z_P \otimes Z_S$ basis, the $Z_P \otimes X_S$ basis, the $X_P \otimes Z_S$ basis and the $X_P \otimes X_S$ basis, hence the former adopts sixteen kinds of initial quantum states; in the latter, Alice generates single photons in two degrees of freedom all in the state of $|R\rangle \otimes |s\rangle$, hence the latter only employs one kind of initial quantum states. We will prove later that by increasing the number of kinds of initial quantum states, the former has higher quantum communication efficiency than the latter; (2) the security check processes of the former are different from those of the latter.

## 4 Security analysis

Firstly, we consider the double CNOT attack from an outside eavesdropper, Eve, which was first suggested by Boyer *et al.* for Eve to attack the mock SQKD protocol of Ref.[2]. Similar to Boyer *et al.*'s secure SQKD protocol in Ref.[2], the proposed protocol can also resist the double CNOT attack from Eve. Concretely speaking, during the transmission of an original single photon with two degrees of freedom from Alice to Bob, Eve may perform the CNOT operation on the original single photon and her auxiliary target photon $|H\rangle \otimes |b_1\rangle$. After that, Eve stores her auxiliary target photon and sends the original single photon to Bob. Bob reflects the original single photon back to Alice or sends a fresh single photon in the same state he found to Alice. In order to make her attack behavior undetected, Eve has to perform the second CNOT operation on the photon from Bob to Alice and her auxiliary target photon. As a result, Eve has no knowledge about the final key bits at all, because the final state of her auxiliary target photon is always $|H\rangle \otimes |b_1\rangle$.

Secondly, we show that the transmissions of single photons between Alice and Bob are completely robust.

When Alice sends single photons to Bob in Step 1, Eve may begin to implement the entangle-measure attack shown as Fig.1. This kind of attack from Eve can be modeled as two unitaries [2-3]: $\hat{U}_E$ attacking the single photons from Alice to Bob and $\hat{U}_F$ attacking the single photons back to Alice, where a common probe space is shared by $\hat{U}_E$ and $\hat{U}_F$.

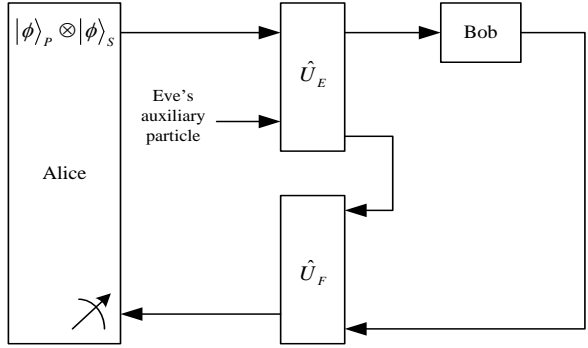

Fig.1 Eve's entangle-measure attack

**Theorem 1.** *Suppose that $\hat{U}_E$ and $\hat{U}_F$ attack the single photon from Alice to Bob and back to Alice, respectively. For no error is caused by this attack in Step 3, the final state of Eve's probe should be irrelevant to Bob's choice of operation and the state in Bob's hand. Hence, Eve obtains nothing useful about the final shared key bits.*

**Proof.** As $|\phi\rangle = |\phi\rangle_P \otimes |\phi\rangle_S$ is within either of the four bases, $Z_P \otimes Z_S$, $Z_P \otimes X_S$, $X_P \otimes Z_S$ and $X_P \otimes X_S$, we prove this theorem according to the following four cases, respectively.

**Case 1:** $|\phi\rangle = |\phi\rangle_P \otimes |\phi\rangle_S$ **is within the** $Z_P \otimes Z_S$ **basis**

(1) Assume that $|\phi\rangle = |\phi\rangle_P \otimes |\phi\rangle_S$ is in the state of $|H\rangle \otimes |b_1\rangle$

The global state of the composite system formed by the single photon $|H\rangle \otimes |b_1\rangle$ and Eve's auxiliary particle $|\varepsilon\rangle$ before Eve's attack can be represented as $(|H\rangle \otimes |b_1\rangle) \otimes |\varepsilon\rangle$. After Eve implements $\hat{U}_E$, the global state is turned into

$$\hat{U}_E ((|H\rangle \otimes |b_1\rangle) \otimes |\varepsilon\rangle) = |Hb_1\rangle |\gamma_{Hb_1}\rangle + |Hb_2\rangle |\gamma_{Hb_2}\rangle + |Vb_1\rangle |\gamma_{Vb_1}\rangle + |Vb_2\rangle |\gamma_{Vb_2}\rangle, \quad (4)$$

where $|\gamma_{Hb_1}\rangle, |\gamma_{Hb_2}\rangle, |\gamma_{Vb_1}\rangle$ and $|\gamma_{Vb_2}\rangle$ are un-normalized states of Eve's probe.

After Bob obtains the state from Alice, he randomly chooses either to CTRL or to SIFT. Eve imposes $\hat{U}_F$ on the state returned to Alice.

i) Consider the situation that Bob chooses to SIFT. Consequently, the global state is collapsed into either of $|Hb_1\rangle |\gamma_{Hb_1}\rangle$, $|Hb_2\rangle |\gamma_{Hb_2}\rangle, |Vb_1\rangle |\gamma_{Vb_1}\rangle$ and $|Vb_2\rangle |\gamma_{Vb_2}\rangle$. For Eve not being discovered in Step 3, it should have

$$\hat{U}_F (|Hb_1\rangle |\gamma_{Hb_1}\rangle) = |Hb_1\rangle |\lambda_{Hb_1}\rangle, \quad (5)$$

$$\hat{U}_F (|Hb_2\rangle |\gamma_{Hb_2}\rangle) = |Hb_2\rangle |\lambda_{Hb_2}\rangle, \quad (6)$$

$$\hat{U}_F(|Vb_1\rangle|\gamma_{Vb_1}\rangle) = |Vb_1\rangle|\lambda_{Vb_1}\rangle, \qquad (7)$$

$$\hat{U}_F(|Vb_2\rangle|\gamma_{Vb_2}\rangle) = |Vb_2\rangle|\lambda_{Vb_2}\rangle, \qquad (8)$$

which imply that $\hat{U}_F$ cannot alter the state of single photon after Bob's measurement, and further

$$|\lambda_{Hb_2}\rangle = |\lambda_{Vb_1}\rangle = |\lambda_{Vb_2}\rangle = 0, \qquad (9)$$

which means that Alice should obtain the measurement results $|Hb_2\rangle$, $|Vb_1\rangle$ and $|Vb_2\rangle$ each with the probability of 0 after her measurement with the $Z_P \otimes Z_S$ basis. Otherwise, Eve can be discovered by Alice with a non-zero probability.

ii) Consider the situation that Bob chooses to CRTL. Consequently, the global state is kept intact.

After Eve imposes $\hat{U}_F$ on the state returned to Alice, due to Eqs.(5-9), the global state is turned into

$$\hat{U}_F(|Hb_1\rangle|\gamma_{Hb_1}\rangle + |Hb_2\rangle|\gamma_{Hb_2}\rangle + |Vb_1\rangle|\gamma_{Vb_1}\rangle + |Vb_2\rangle|\gamma_{Vb_2}\rangle) = |Hb_1\rangle|\lambda_{Hb_1}\rangle = |Hb_1\rangle|\lambda\rangle. \qquad (10)$$

For Eve not being discovered in Step 3, Alice's measurement result should be $|H\rangle \otimes |b_1\rangle$. Apparently, Eq.(10) automatically meets this requirement.

It can be concluded from Eq.(5) and Eq.(10) that, in this circumstance, for Eve not inducing errors in Step 3, the final state of Eve's probe should be independent of Bob's choice of operation.

(2) Assume that $|\phi\rangle = |\phi\rangle_P \otimes |\phi\rangle_S$ is in the state of $|H\rangle \otimes |b_2\rangle$

The global state of the composite system formed by the single photon $|H\rangle \otimes |b_2\rangle$ and Eve's auxiliary particle $|\varepsilon\rangle$ before Eve's attack can be represented as $(|H\rangle \otimes |b_2\rangle) \otimes |\varepsilon\rangle$. After Eve implements $\hat{U}_E$, the global state is turned into

$$\hat{U}_E((|H\rangle \otimes |b_2\rangle) \otimes |\varepsilon\rangle) = |Hb_1\rangle|\mu_{Hb_1}\rangle + |Hb_2\rangle|\mu_{Hb_2}\rangle + |Vb_1\rangle|\mu_{Vb_1}\rangle + |Vb_2\rangle|\mu_{Vb_2}\rangle, \qquad (11)$$

where $|\mu_{Hb_1}\rangle$, $|\mu_{Hb_2}\rangle$, $|\mu_{Vb_1}\rangle$ and $|\mu_{Vb_2}\rangle$ are un-normalized states of Eve's probe.

After Bob obtains the state from Alice, he randomly chooses either to CTRL or to SIFT. Eve imposes $\hat{U}_F$ on the state returned to Alice.

i) Consider the situation that Bob chooses to SIFT. Consequently, the global state is collapsed into either of $|Hb_1\rangle|\mu_{Hb_1}\rangle$, $|Hb_2\rangle|\mu_{Hb_2}\rangle$, $|Vb_1\rangle|\mu_{Vb_1}\rangle$ and $|Vb_2\rangle|\mu_{Vb_2}\rangle$. For Eve not being discovered in Step 3, it should have

$$\hat{U}_F(|Hb_1\rangle|\mu_{Hb_1}\rangle) = |Hb_1\rangle|\nu_{Hb_1}\rangle, \qquad (12)$$

$$\hat{U}_F(|Hb_2\rangle|\mu_{Hb_2}\rangle) = |Hb_2\rangle|\nu_{Hb_2}\rangle, \qquad (13)$$

$$\hat{U}_F(|Vb_1\rangle|\mu_{Vb_1}\rangle) = |Vb_1\rangle|\nu_{Vb_1}\rangle, \qquad (14)$$

$$\hat{U}_F(|Vb_2\rangle|\mu_{Vb_2}\rangle) = |Vb_2\rangle|\nu_{Vb_2}\rangle, \qquad (15)$$

which imply that $\hat{U}_F$ cannot alter the state of single photon after Bob's measurement, and further

$$|\nu_{Hb_1}\rangle = |\nu_{Vb_1}\rangle = |\nu_{Vb_2}\rangle = 0, \qquad (16)$$

which means that Alice should obtain the measurement results $|Hb_1\rangle$, $|Vb_1\rangle$ and $|Vb_2\rangle$ each with the probability of 0 after her measurement with the $Z_P \otimes Z_S$ basis. Otherwise, Eve can be discovered by Alice with a non-zero probability.

ii) Consider the situation that Bob chooses to CRTL. Consequently, the global state is kept intact.

After Eve imposes $\hat{U}_F$ on the state returned to Alice, due to Eqs.(12-16), the global state is turned into

$$\hat{U}_F(|Hb_1\rangle|\mu_{Hb_1}\rangle + |Hb_2\rangle|\mu_{Hb_2}\rangle + |Vb_1\rangle|\mu_{Vb_1}\rangle + |Vb_2\rangle|\mu_{Vb_2}\rangle) = |Hb_2\rangle|\nu_{Hb_2}\rangle = |Hb_2\rangle|\nu\rangle. \qquad (17)$$

For Eve not being discovered in Step 3, Alice's measurement result should be $|H\rangle \otimes |b_2\rangle$. Apparently, Eq.(17) automatically meets this requirement.

It can be concluded from Eq.(13) and Eq.(17) that, in this circumstance, for Eve not inducing errors in Step 3, the final state of Eve's probe should be independent of Bob's choice of operation.

(3) Assume that $|\phi\rangle = |\phi\rangle_P \otimes |\phi\rangle_S$ is in the state of $|V\rangle \otimes |b_1\rangle$

The global state of the composite system formed by the single photon $|V\rangle \otimes |b_1\rangle$ and Eve's auxiliary particle $|\varepsilon\rangle$ before Eve's attack can be represented as $(|V\rangle \otimes |b_1\rangle) \otimes |\varepsilon\rangle$. After Eve implements $\hat{U}_E$, the global state is turned into

$$\hat{U}_E((|V\rangle \otimes |b_1\rangle) \otimes |\varepsilon\rangle) = |Hb_1\rangle|\theta_{Hb_1}\rangle + |Hb_2\rangle|\theta_{Hb_2}\rangle + |Vb_1\rangle|\theta_{Vb_1}\rangle + |Vb_2\rangle|\theta_{Vb_2}\rangle, \qquad (18)$$

where $|\theta_{Hb_1}\rangle, |\theta_{Hb_2}\rangle, |\theta_{Vb_1}\rangle$ and $|\theta_{Vb_2}\rangle$ are un-normalized states of Eve's probe.

After Bob obtains the state from Alice, he randomly chooses either to CTRL or to SIFT. Eve imposes $\hat{U}_F$ on the state returned to Alice.

i) Consider the situation that Bob chooses to SIFT. Consequently, the global state is collapsed into either of $|Hb_1\rangle|\theta_{Hb_1}\rangle$, $|Hb_2\rangle|\theta_{Hb_2}\rangle$, $|Vb_1\rangle|\theta_{Vb_1}\rangle$ and $|Vb_2\rangle|\theta_{Vb_2}\rangle$. For Eve not being discovered in Step 3, it should have

$$\hat{U}_F(|Hb_1\rangle|\theta_{Hb_1}\rangle) = |Hb_1\rangle|\vartheta_{Hb_1}\rangle, \qquad (19)$$

$$\hat{U}_F(|Hb_2\rangle|\theta_{Hb_2}\rangle) = |Hb_2\rangle|\vartheta_{Hb_2}\rangle, \qquad (20)$$

$$\hat{U}_F(|Vb_1\rangle|\theta_{Vb_1}\rangle) = |Vb_1\rangle|\vartheta_{Vb_1}\rangle, \qquad (21)$$

$$\hat{U}_F(|Vb_2\rangle|\theta_{Vb_2}\rangle) = |Vb_2\rangle|\vartheta_{Vb_2}\rangle, \qquad (22)$$

which imply that $\hat{U}_F$ cannot alter the state of single photon after Bob's measurement, and further

$$|\vartheta_{Hb_1}\rangle = |\vartheta_{Hb_2}\rangle = |\vartheta_{Vb_2}\rangle = 0, \qquad (23)$$

which means that Alice should obtain the measurement results $|Hb_1\rangle$, $|Hb_2\rangle$ and $|Vb_2\rangle$ each with the probability of 0 after her measurement with the $Z_P \otimes Z_S$ basis. Otherwise, Eve can be discovered by Alice with a non-zero probability.

ii) Consider the situation that Bob chooses to CRTL. Consequently, the global state is kept intact.

After Eve imposes $\hat{U}_F$ on the state returned to Alice, due to Eqs.(19-23), the global state is turned into

$$\hat{U}_F(|Hb_1\rangle|\theta_{Hb_1}\rangle + |Hb_2\rangle|\theta_{Hb_2}\rangle + |Vb_1\rangle|\theta_{Vb_1}\rangle + |Vb_2\rangle|\theta_{Vb_2}\rangle) = |Vb_1\rangle|\vartheta_{Vb_1}\rangle = |Vb_1\rangle|\vartheta\rangle. \qquad (24)$$

For Eve not being discovered in Step 3, Alice's measurement result should be $|V\rangle \otimes |b_1\rangle$. Apparently, Eq.(24) automatically meets this requirement.

It can be concluded from Eq.(21) and Eq.(24) that, in this circumstance, for Eve not inducing errors in Step 3, the final state of Eve's probe should be independent of Bob's choice of operation.

(4) Assume that $|\phi\rangle = |\phi\rangle_P \otimes |\phi\rangle_S$ is in the state of $|V\rangle \otimes |b_2\rangle$

The global state of the composite system formed by the single photon $|V\rangle \otimes |b_2\rangle$ and Eve's auxiliary particle $|\varepsilon\rangle$ before Eve's attack can be represented as $(|V\rangle \otimes |b_2\rangle) \otimes |\varepsilon\rangle$. After Eve implements $\hat{U}_E$, the global state is turned into

$$\hat{U}_E((|V\rangle \otimes |b_2\rangle) \otimes |\varepsilon\rangle) = |Hb_1\rangle|\sigma_{Hb_1}\rangle + |Hb_2\rangle|\sigma_{Hb_2}\rangle + |Vb_1\rangle|\sigma_{Vb_1}\rangle + |Vb_2\rangle|\sigma_{Vb_2}\rangle, \qquad (25)$$

where $|\sigma_{Hb_1}\rangle, |\sigma_{Hb_2}\rangle, |\sigma_{Vb_1}\rangle$ and $|\sigma_{Vb_2}\rangle$ are un-normalized states of Eve's probe.

After Bob obtains the state from Alice, he randomly chooses either to CTRL or to SIFT. Eve imposes $\hat{U}_F$ on the state returned to Alice.

i) Consider the situation that Bob chooses to SIFT. Consequently, the global state is collapsed into either of $|Hb_1\rangle|\sigma_{Hb_1}\rangle$, $|Hb_2\rangle|\sigma_{Hb_2}\rangle$, $|Vb_1\rangle|\sigma_{Vb_1}\rangle$ and $|Vb_2\rangle|\sigma_{Vb_2}\rangle$. For Eve not being discovered in Step 3, it should have

$$\hat{U}_F(|Hb_1\rangle|\sigma_{Hb_1}\rangle) = |Hb_1\rangle|\tau_{Hb_1}\rangle, \qquad (26)$$

$$\hat{U}_F(|Hb_2\rangle|\sigma_{Hb_2}\rangle) = |Hb_2\rangle|\tau_{Hb_2}\rangle, \qquad (27)$$

$$\hat{U}_F(|Vb_1\rangle|\sigma_{Vb_1}\rangle) = |Vb_1\rangle|\tau_{Vb_1}\rangle, \qquad (28)$$

$$\hat{U}_F(|Vb_2\rangle|\sigma_{Vb_2}\rangle) = |Vb_2\rangle|\tau_{Vb_2}\rangle, \qquad (29)$$

which imply that $\hat{U}_F$ cannot alter the state of single photon after Bob's measurement, and further

$$|\tau_{Hb_1}\rangle = |\tau_{Hb_2}\rangle = |\tau_{Vb_1}\rangle = 0, \qquad (30)$$

which means that Alice should obtain the measurement results $|Hb_1\rangle$, $|Hb_2\rangle$ and $|Vb_1\rangle$ each with the probability of 0 after her measurement with the $Z_P \otimes Z_S$ basis. Otherwise, Eve can be discovered by Alice with a non-zero probability.

ii) Consider the situation that Bob chooses to CRTL. Consequently, the global state is kept intact.

After Eve imposes $\hat{U}_F$ on the state returned to Alice, due to Eqs.(26-30), the global state is turned into

$$\hat{U}_F \left( |Hb_1\rangle|\sigma_{Hb_1}\rangle + |Hb_2\rangle|\sigma_{Hb_2}\rangle + |Vb_1\rangle|\sigma_{Vb_1}\rangle + |Vb_2\rangle|\sigma_{Vb_2}\rangle \right) = |Vb_2\rangle|\tau_{Vb_2}\rangle = |Vb_2\rangle|\tau\rangle. \quad (31)$$

For Eve not being discovered in Step 3, Alice's measurement result should be $|V\rangle \otimes |b_2\rangle$. Apparently, Eq.(31) automatically meets this requirement.

It can be concluded from Eq.(29) and Eq.(31) that, in this circumstance, for Eve not inducing errors in Step 3, the final state of Eve's probe should be independent of Bob's choice of operation.

**Case 2:** $|\phi\rangle = |\phi\rangle_P \otimes |\phi\rangle_S$ **is within the** $Z_P \otimes X_S$ **basis**

(1) Assume that $|\phi\rangle = |\phi\rangle_P \otimes |\phi\rangle_S$ is in the state of $|H\rangle \otimes |s\rangle$

The global state of the composite system formed by the single photon $|H\rangle \otimes |s\rangle$ and Eve's auxiliary particle $|\varepsilon\rangle$ before Eve's attack can be represented as $(|H\rangle \otimes |s\rangle) \otimes |\varepsilon\rangle$. According to the linearity of quantum mechanics together with Eq.(4) and Eq.(11), after Eve implements $\hat{U}_E$, the global state is turned into

$$\hat{U}_E\left((|H\rangle \otimes |s\rangle) \otimes |\varepsilon\rangle\right) = \hat{U}_E\left(\left(|H\rangle \otimes \frac{1}{\sqrt{2}}(|b_1\rangle + |b_2\rangle)\right) \otimes |\varepsilon\rangle\right)$$

$$= \frac{1}{\sqrt{2}}\left(|Hb_1\rangle|\gamma_{Hb_1}\rangle + |Hb_2\rangle|\gamma_{Hb_2}\rangle + |Vb_1\rangle|\gamma_{Vb_1}\rangle + |Vb_2\rangle|\gamma_{Vb_2}\rangle\right)$$

$$+ \frac{1}{\sqrt{2}}\left(|Hb_1\rangle|\mu_{Hb_1}\rangle + |Hb_2\rangle|\mu_{Hb_2}\rangle + |Vb_1\rangle|\mu_{Vb_1}\rangle + |Vb_2\rangle|\mu_{Vb_2}\rangle\right). \quad (32)$$

After Bob obtains the state from Alice, he randomly chooses either to CTRL or to SIFT. Eve imposes $\hat{U}_F$ on the state returned to Alice.

i) Consider the situation that Bob chooses to CRTL. Consequently, the global state is kept intact.

After Eve imposes $\hat{U}_F$ on the state returned to Alice, due to the linearity of quantum mechanics together with Eq.(10) and Eq.(17), the global state of Eq.(32) is turned into

$$\hat{U}_F\left(\hat{U}_E\left((|H\rangle \otimes |s\rangle) \otimes |\varepsilon\rangle\right)\right) = \frac{1}{\sqrt{2}}\hat{U}_F\left(|Hb_1\rangle|\gamma_{Hb_1}\rangle + |Hb_2\rangle|\gamma_{Hb_2}\rangle + |Vb_1\rangle|\gamma_{Vb_1}\rangle + |Vb_2\rangle|\gamma_{Vb_2}\rangle\right)$$

$$+ \frac{1}{\sqrt{2}}\hat{U}_F\left(|Hb_1\rangle|\mu_{Hb_1}\rangle + |Hb_2\rangle|\mu_{Hb_2}\rangle + |Vb_1\rangle|\mu_{Vb_1}\rangle + |Vb_2\rangle|\mu_{Vb_2}\rangle\right)$$

$$= \frac{1}{\sqrt{2}}|Hb_1\rangle|\lambda\rangle + \frac{1}{\sqrt{2}}|Hb_2\rangle|\nu\rangle. \quad (33)$$

For Eve not being discovered in Step 3, Alice's measurement result should be $|H\rangle \otimes |s\rangle$. Thus, it can be obtained from Eq.(33) that

$$|\lambda\rangle = |\nu\rangle. \quad (34)$$

ii) Consider the situation that Bob chooses to SIFT. Consequently, the global state is collapsed into either of $|Hb_1\rangle|\gamma_{Hb_1}\rangle$, $|Hb_2\rangle|\gamma_{Hb_2}\rangle$, $|Vb_1\rangle|\gamma_{Vb_1}\rangle$, $|Vb_2\rangle|\gamma_{Vb_2}\rangle$, $|Hb_1\rangle|\mu_{Hb_1}\rangle$, $|Hb_2\rangle|\mu_{Hb_2}\rangle$, $|Vb_1\rangle|\mu_{Vb_1}\rangle$ and $|Vb_2\rangle|\mu_{Vb_2}\rangle$. According to Eqs.(5-8) and Eqs.(12-15), $\hat{U}_F$ automatically keeps the state of single photon after Bob's measurement unchanged. Further, according to Eqs.(9-10), Eqs.(16-17) and Eq.(34), Alice can only randomly obtain the measurement results $|Hb_1\rangle$ and $|Hb_2\rangle$ after her measurement with the $Z_P \otimes Z_S$ basis. Hence, in this situation, Eve is not detectable in Step 3.

It can be concluded that, in this circumstance, for Eve not inducing errors in Step 3, the final state of Eve's probe should be independent of Bob's choice of operation.

(2) Assume that $|\phi\rangle = |\phi\rangle_P \otimes |\phi\rangle_S$ is in the state of $|H\rangle \otimes |a\rangle$

The global state of the composite system formed by the single photon $|H\rangle\otimes|a\rangle$ and Eve's auxiliary particle $|\varepsilon\rangle$ before Eve's attack can be represented as $(|H\rangle\otimes|a\rangle)\otimes|\varepsilon\rangle$. According to the linearity of quantum mechanics together with Eq.(4) and Eq.(11), after Eve implements $\hat{U}_E$, the global state is turned into

$$\hat{U}_E((|H\rangle\otimes|a\rangle)\otimes|\varepsilon\rangle) = \hat{U}_E\left(\left(|H\rangle\otimes\frac{1}{\sqrt{2}}(|b_1\rangle-|b_2\rangle)\right)\otimes|\varepsilon\rangle\right)$$

$$= \frac{1}{\sqrt{2}}(|Hb_1\rangle|\gamma_{Hb_1}\rangle+|Hb_2\rangle|\gamma_{Hb_2}\rangle+|Vb_1\rangle|\gamma_{Vb_1}\rangle+|Vb_2\rangle|\gamma_{Vb_2}\rangle)$$

$$-\frac{1}{\sqrt{2}}(|Hb_1\rangle|\mu_{Hb_1}\rangle+|Hb_2\rangle|\mu_{Hb_2}\rangle+|Vb_1\rangle|\mu_{Vb_1}\rangle+|Vb_2\rangle|\mu_{Vb_2}\rangle). \quad (35)$$

After Bob obtains the state from Alice, he randomly chooses either to CTRL or to SIFT. Eve imposes $\hat{U}_F$ on the state returned to Alice.

i) Consider the situation that Bob chooses to CRTL. Consequently, the global state is kept intact.

After Eve imposes $\hat{U}_F$ on the state returned to Alice, due to the linearity of quantum mechanics together with Eq.(10) and Eq.(17), the global state of Eq.(35) is turned into

$$\hat{U}_F\left(\hat{U}_E((|H\rangle\otimes|a\rangle)\otimes|\varepsilon\rangle)\right) = \frac{1}{\sqrt{2}}\hat{U}_F(|Hb_1\rangle|\gamma_{Hb_1}\rangle+|Hb_2\rangle|\gamma_{Hb_2}\rangle+|Vb_1\rangle|\gamma_{Vb_1}\rangle+|Vb_2\rangle|\gamma_{Vb_2}\rangle)$$

$$-\frac{1}{\sqrt{2}}\hat{U}_F(|Hb_1\rangle|\mu_{Hb_1}\rangle+|Hb_2\rangle|\mu_{Hb_2}\rangle+|Vb_1\rangle|\mu_{Vb_1}\rangle+|Vb_2\rangle|\mu_{Vb_2}\rangle)$$

$$= \frac{1}{\sqrt{2}}|Hb_1\rangle|\lambda\rangle - \frac{1}{\sqrt{2}}|Hb_2\rangle|\nu\rangle. \quad (36)$$

For Eve not being discovered in Step 3, Alice's measurement result should be $|H\rangle\otimes|a\rangle$. It is naturally derived after Eq.(34) is inserted into Eq.(36).

ii) Consider the situation that Bob chooses to SIFT. Consequently, the global state is collapsed into either of $|Hb_1\rangle|\gamma_{Hb_1}\rangle$, $|Hb_2\rangle|\gamma_{Hb_2}\rangle$, $|Vb_1\rangle|\gamma_{Vb_1}\rangle$, $|Vb_2\rangle|\gamma_{Vb_2}\rangle$, $|Hb_1\rangle|\mu_{Hb_1}\rangle$, $|Hb_2\rangle|\mu_{Hb_2}\rangle$, $|Vb_1\rangle|\mu_{Vb_1}\rangle$ and $|Vb_2\rangle|\mu_{Vb_2}\rangle$. According to Eqs.(5-8) and Eqs.(12-15), $\hat{U}_F$ automatically keeps the state of single photon after Bob's measurement unchanged. Further, according to Eqs.(9-10), Eqs.(16-17) and Eq.(34), Alice only can randomly obtain the measurement results $|Hb_1\rangle$ and $|Hb_2\rangle$ after her measurement with the $Z_P\otimes Z_S$ basis. Hence, in this situation, Eve is not detectable in Step 3.

It can be concluded that, in this circumstance, for Eve not inducing errors in Step 3, the final state of Eve's probe should be independent of Bob's choice of operation.

(3) Assume that $|\phi\rangle = |\phi\rangle_P\otimes|\phi\rangle_S$ is in the state of $|V\rangle\otimes|s\rangle$

The global state of the composite system formed by the single photon $|V\rangle\otimes|s\rangle$ and Eve's auxiliary particle $|\varepsilon\rangle$ before Eve's attack can be represented as $(|V\rangle\otimes|s\rangle)\otimes|\varepsilon\rangle$. According to the linearity of quantum mechanics together with Eq.(18) and Eq.(25), after Eve implements $\hat{U}_E$, the global state is turned into

$$\hat{U}_E((|V\rangle\otimes|s\rangle)\otimes|\varepsilon\rangle) = \hat{U}_E\left(\left(|V\rangle\otimes\frac{1}{\sqrt{2}}(|b_1\rangle+|b_2\rangle)\right)\otimes|\varepsilon\rangle\right)$$

$$= \frac{1}{\sqrt{2}}(|Hb_1\rangle|\theta_{Hb_1}\rangle+|Hb_2\rangle|\theta_{Hb_2}\rangle+|Vb_1\rangle|\theta_{Vb_1}\rangle+|Vb_2\rangle|\theta_{Vb_2}\rangle)$$

$$+\frac{1}{\sqrt{2}}(|Hb_1\rangle|\sigma_{Hb_1}\rangle+|Hb_2\rangle|\sigma_{Hb_2}\rangle+|Vb_1\rangle|\sigma_{Vb_1}\rangle+|Vb_2\rangle|\sigma_{Vb_2}\rangle). \quad (37)$$

After Bob obtains the state from Alice, he randomly chooses either to CTRL or to SIFT. Eve imposes $\hat{U}_F$ on the state returned to Alice.

i) Consider the situation that Bob chooses to CRTL. Consequently, the global state is kept intact.

After Eve imposes $\hat{U}_F$ on the state returned to Alice, due to the linearity of quantum mechanics together with Eq.(24) and Eq.(31), the global state of Eq.(37) is turned into

$$\hat{U}_F\left(\hat{U}_E\left(\left(|V\rangle\otimes|s\rangle\right)\otimes|\varepsilon\rangle\right)\right) = \frac{1}{\sqrt{2}}\hat{U}_F\left(|Hb_1\rangle|\theta_{Hb_1}\rangle + |Hb_2\rangle|\theta_{Hb_2}\rangle + |Vb_1\rangle|\theta_{Vb_1}\rangle + |Vb_2\rangle|\theta_{Vb_2}\rangle\right)$$
$$+ \frac{1}{\sqrt{2}}\hat{U}_F\left(|Hb_1\rangle|\sigma_{Hb_1}\rangle + |Hb_2\rangle|\sigma_{Hb_2}\rangle + |Vb_1\rangle|\sigma_{Vb_1}\rangle + |Vb_2\rangle|\sigma_{Vb_2}\rangle\right)$$
$$= \frac{1}{\sqrt{2}}|Vb_1\rangle|\vartheta\rangle + \frac{1}{\sqrt{2}}|Vb_2\rangle|\tau\rangle. \quad (38)$$

For Eve not being discovered in Step 3, Alice's measurement result should be $|V\rangle\otimes|s\rangle$. Thus, it can be obtained from Eq.(38) that

$$|\vartheta\rangle = |\tau\rangle. \quad (39)$$

ii) Consider the situation that Bob chooses to SIFT. Consequently, the global state is collapsed into either of $|Hb_1\rangle|\theta_{Hb_1}\rangle$, $|Hb_2\rangle|\theta_{Hb_2}\rangle$, $|Vb_1\rangle|\theta_{Vb_1}\rangle$, $|Vb_2\rangle|\theta_{Vb_2}\rangle$, $|Hb_1\rangle|\sigma_{Hb_1}\rangle$, $|Hb_2\rangle|\sigma_{Hb_2}\rangle$, $|Vb_1\rangle|\sigma_{Vb_1}\rangle$ and $|Vb_2\rangle|\sigma_{Vb_2}\rangle$. According to Eqs.(19-22) and Eqs.(26-29), $\hat{U}_F$ automatically keeps the state of single photon after Bob's measurement unchanged. Further, according to Eqs.(23-24), Eqs.(30-31) and Eq.(39), Alice only can randomly obtain the measurement results $|Vb_1\rangle$ and $|Vb_2\rangle$ after her measurement with the $Z_P\otimes Z_S$ basis. Hence, in this situation, Eve is not detectable in Step 3.

It can be concluded that, in this circumstance, for Eve not inducing errors in Step 3, the final state of Eve's probe should be independent of Bob's choice of operation.

(4) Assume that $|\phi\rangle = |\phi\rangle_P \otimes |\phi\rangle_S$ is in the state of $|V\rangle\otimes|a\rangle$

The global state of the composite system formed by the single photon $|V\rangle\otimes|a\rangle$ and Eve's auxiliary particle $|\varepsilon\rangle$ before Eve's attack can be represented as $\left(|V\rangle\otimes|a\rangle\right)\otimes|\varepsilon\rangle$. According to the linearity of quantum mechanics together with Eq.(18) and Eq.(25), after Eve implements $\hat{U}_E$, the global state is turned into

$$\hat{U}_E\left(\left(|V\rangle\otimes|a\rangle\right)\otimes|\varepsilon\rangle\right) = \hat{U}_E\left(\left(|V\rangle\otimes\frac{1}{\sqrt{2}}\left(|b_1\rangle - |b_2\rangle\right)\right)\otimes|\varepsilon\rangle\right)$$
$$= \frac{1}{\sqrt{2}}\left(|Hb_1\rangle|\theta_{Hb_1}\rangle + |Hb_2\rangle|\theta_{Hb_2}\rangle + |Vb_1\rangle|\theta_{Vb_1}\rangle + |Vb_2\rangle|\theta_{Vb_2}\rangle\right)$$
$$- \frac{1}{\sqrt{2}}\left(|Hb_1\rangle|\sigma_{Hb_1}\rangle + |Hb_2\rangle|\sigma_{Hb_2}\rangle + |Vb_1\rangle|\sigma_{Vb_1}\rangle + |Vb_2\rangle|\sigma_{Vb_2}\rangle\right). \quad (40)$$

After Bob obtains the state from Alice, he randomly chooses either to CTRL or to SIFT. Eve imposes $\hat{U}_F$ on the state returned to Alice.

i) Consider the situation that Bob chooses to CRTL. Consequently, the global state is kept intact.

After Eve imposes $\hat{U}_F$ on the state returned to Alice, due to the linearity of quantum mechanics together with Eq.(24) and Eq.(31), the global state of Eq.(40) is turned into

$$\hat{U}_F\left(\hat{U}_E\left(\left(|V\rangle\otimes|a\rangle\right)\otimes|\varepsilon\rangle\right)\right) = \frac{1}{\sqrt{2}}\hat{U}_F\left(|Hb_1\rangle|\theta_{Hb_1}\rangle + |Hb_2\rangle|\theta_{Hb_2}\rangle + |Vb_1\rangle|\theta_{Vb_1}\rangle + |Vb_2\rangle|\theta_{Vb_2}\rangle\right)$$
$$- \frac{1}{\sqrt{2}}\hat{U}_F\left(|Hb_1\rangle|\sigma_{Hb_1}\rangle + |Hb_2\rangle|\sigma_{Hb_2}\rangle + |Vb_1\rangle|\sigma_{Vb_1}\rangle + |Vb_2\rangle|\sigma_{Vb_2}\rangle\right)$$
$$= \frac{1}{\sqrt{2}}|Vb_1\rangle|\vartheta\rangle - \frac{1}{\sqrt{2}}|Vb_2\rangle|\tau\rangle. \quad (41)$$

For Eve not being discovered in Step 3, Alice's measurement result should be $|V\rangle\otimes|a\rangle$. It is naturally derived after Eq.(39) is inserted into Eq.(41).

ii) Consider the situation that Bob chooses to SIFT. Consequently, the global state is collapsed into either of $|Hb_1\rangle|\theta_{Hb_1}\rangle$, $|Hb_2\rangle|\theta_{Hb_2}\rangle$, $|Vb_1\rangle|\theta_{Vb_1}\rangle$, $|Vb_2\rangle|\theta_{Vb_2}\rangle$, $|Hb_1\rangle|\sigma_{Hb_1}\rangle$, $|Hb_2\rangle|\sigma_{Hb_2}\rangle$, $|Vb_1\rangle|\sigma_{Vb_1}\rangle$ and $|Vb_2\rangle|\sigma_{Vb_2}\rangle$. According to Eqs.(19-22) and Eqs.(26-29), $\hat{U}_F$ automatically keeps the state of single photon after Bob's measurement unchanged. Further, according to Eqs.(23-24), Eqs.(30-31) and Eq.(39), Alice only can randomly obtain the measurement results $|Vb_1\rangle$ and $|Vb_2\rangle$ after her measurement with the $Z_P \otimes Z_S$ basis. Hence, in this situation, Eve is not detectable in Step 3.

It can be concluded that, in this circumstance, for Eve not inducing errors in Step 3, the final state of Eve's probe should be independent of Bob's choice of operation.

**Case 3:** $|\phi\rangle = |\phi\rangle_P \otimes |\phi\rangle_S$ **is within the** $X_P \otimes Z_S$ **basis**

(1) Assume that $|\phi\rangle = |\phi\rangle_P \otimes |\phi\rangle_S$ is in the state of $|R\rangle \otimes |b_1\rangle$

The global state of the composite system formed by the single photon $|R\rangle \otimes |b_1\rangle$ and Eve's auxiliary particle $|\varepsilon\rangle$ before Eve's attack can be represented as $(|R\rangle \otimes |b_1\rangle) \otimes |\varepsilon\rangle$. According to the linearity of quantum mechanics together with Eq.(4) and Eq.(18), after Eve implements $\hat{U}_E$, the global state is turned into

$$\hat{U}_E((|R\rangle \otimes |b_1\rangle) \otimes |\varepsilon\rangle) = \hat{U}_E\left(\left(\frac{1}{\sqrt{2}}(|H\rangle + |V\rangle) \otimes |b_1\rangle\right) \otimes |\varepsilon\rangle\right)$$
$$= \frac{1}{\sqrt{2}}(|Hb_1\rangle|\gamma_{Hb_1}\rangle + |Hb_2\rangle|\gamma_{Hb_2}\rangle + |Vb_1\rangle|\gamma_{Vb_1}\rangle + |Vb_2\rangle|\gamma_{Vb_2}\rangle)$$
$$+ \frac{1}{\sqrt{2}}(|Hb_1\rangle|\theta_{Hb_1}\rangle + |Hb_2\rangle|\theta_{Hb_2}\rangle + |Vb_1\rangle|\theta_{Vb_1}\rangle + |Vb_2\rangle|\theta_{Vb_2}\rangle). \quad (42)$$

After Bob obtains the state from Alice, he randomly chooses either to CTRL or to SIFT. Eve imposes $\hat{U}_F$ on the state returned to Alice.

i) Consider the situation that Bob chooses to CRTL. Consequently, the global state is kept intact.

After Eve imposes $\hat{U}_F$ on the state returned to Alice, due to the linearity of quantum mechanics together with Eq.(10) and Eq.(24), the global state of Eq.(42) is turned into

$$\hat{U}_F\left(\hat{U}_E((|R\rangle \otimes |b_1\rangle) \otimes |\varepsilon\rangle)\right) = \frac{1}{\sqrt{2}}\hat{U}_F(|Hb_1\rangle|\gamma_{Hb_1}\rangle + |Hb_2\rangle|\gamma_{Hb_2}\rangle + |Vb_1\rangle|\gamma_{Vb_1}\rangle + |Vb_2\rangle|\gamma_{Vb_2}\rangle)$$
$$+ \frac{1}{\sqrt{2}}\hat{U}_F(|Hb_1\rangle|\theta_{Hb_1}\rangle + |Hb_2\rangle|\theta_{Hb_2}\rangle + |Vb_1\rangle|\theta_{Vb_1}\rangle + |Vb_2\rangle|\theta_{Vb_2}\rangle)$$
$$= \frac{1}{\sqrt{2}}|Hb_1\rangle|\lambda\rangle + \frac{1}{\sqrt{2}}|Vb_1\rangle|\vartheta\rangle. \quad (43)$$

For Eve not being discovered in Step 3, Alice's measurement result should be $|R\rangle \otimes |b_1\rangle$. Thus, it can be obtained from Eq.(43) that

$$|\lambda\rangle = |\vartheta\rangle. \quad (44)$$

ii) Consider the situation that Bob chooses to SIFT. Consequently, the global state is collapsed into either of $|Hb_1\rangle|\gamma_{Hb_1}\rangle$, $|Hb_2\rangle|\gamma_{Hb_2}\rangle$, $|Vb_1\rangle|\gamma_{Vb_1}\rangle$, $|Vb_2\rangle|\gamma_{Vb_2}\rangle$, $|Hb_1\rangle|\theta_{Hb_1}\rangle$, $|Hb_2\rangle|\theta_{Hb_2}\rangle$, $|Vb_1\rangle|\theta_{Vb_1}\rangle$ and $|Vb_2\rangle|\theta_{Vb_2}\rangle$. According to Eqs.(5-8) and Eqs.(19-22), $\hat{U}_F$ automatically keeps the state of single photon after Bob's measurement unchanged. Further, according to Eqs.(9-10), Eqs.(23-24) and Eq.(44), Alice only can randomly obtain the measurement results $|Hb_1\rangle$ and $|Vb_1\rangle$ after her measurement with the $Z_P \otimes Z_S$ basis. Hence, in this situation, Eve is not detectable in Step 3.

It can be concluded that, in this circumstance, for Eve not inducing errors in Step 3, the final state of Eve's probe should be independent of Bob's choice of operation.

(2) Assume that $|\phi\rangle = |\phi\rangle_P \otimes |\phi\rangle_S$ is in the state of $|A\rangle \otimes |b_1\rangle$

The global state of the composite system formed by the single photon $|A\rangle \otimes |b_1\rangle$ and Eve's auxiliary particle $|\varepsilon\rangle$ before Eve's attack can be represented as $(|A\rangle \otimes |b_1\rangle) \otimes |\varepsilon\rangle$. According to the linearity of quantum mechanics together with Eq.(4) and Eq.(18), after Eve implements $\hat{U}_E$, the global state is turned into

$$\hat{U}_E((|A\rangle \otimes |b_1\rangle) \otimes |\varepsilon\rangle) = \hat{U}_E\left(\left(\frac{1}{\sqrt{2}}(|H\rangle - |V\rangle) \otimes |b_1\rangle\right) \otimes |\varepsilon\rangle\right)$$

$$= \frac{1}{\sqrt{2}}(|Hb_1\rangle|\gamma_{Hb_1}\rangle + |Hb_2\rangle|\gamma_{Hb_2}\rangle + |Vb_1\rangle|\gamma_{Vb_1}\rangle + |Vb_2\rangle|\gamma_{Vb_2}\rangle)$$

$$- \frac{1}{\sqrt{2}}(|Hb_1\rangle|\theta_{Hb_1}\rangle + |Hb_2\rangle|\theta_{Hb_2}\rangle + |Vb_1\rangle|\theta_{Vb_1}\rangle + |Vb_2\rangle|\theta_{Vb_2}\rangle). \tag{45}$$

After Bob obtains the state from Alice, he randomly chooses either to CTRL or to SIFT. Eve imposes $\hat{U}_F$ on the state returned to Alice.

i) Consider the situation that Bob chooses to CRTL. Consequently, the global state is kept intact.

After Eve imposes $\hat{U}_F$ on the state returned to Alice, due to the linearity of quantum mechanics together with Eq.(10) and Eq.(24), the global state of Eq.(45) is turned into

$$\hat{U}_F\left(\hat{U}_E((|A\rangle \otimes |b_1\rangle) \otimes |\varepsilon\rangle)\right) = \frac{1}{\sqrt{2}} \hat{U}_F(|Hb_1\rangle|\gamma_{Hb_1}\rangle + |Hb_2\rangle|\gamma_{Hb_2}\rangle + |Vb_1\rangle|\gamma_{Vb_1}\rangle + |Vb_2\rangle|\gamma_{Vb_2}\rangle)$$

$$- \frac{1}{\sqrt{2}} \hat{U}_F(|Hb_1\rangle|\theta_{Hb_1}\rangle + |Hb_2\rangle|\theta_{Hb_2}\rangle + |Vb_1\rangle|\theta_{Vb_1}\rangle + |Vb_2\rangle|\theta_{Vb_2}\rangle)$$

$$= \frac{1}{\sqrt{2}}|Hb_1\rangle|\lambda\rangle - \frac{1}{\sqrt{2}}|Vb_1\rangle|\vartheta\rangle. \tag{46}$$

For Eve not being discovered in Step 3, Alice's measurement result should be $|A\rangle \otimes |b_1\rangle$. It is naturally derived after Eq.(44) is inserted into Eq.(46).

ii) Consider the situation that Bob chooses to SIFT. Consequently, the global state is collapsed into either of $|Hb_1\rangle|\gamma_{Hb_1}\rangle$, $|Hb_2\rangle|\gamma_{Hb_2}\rangle$, $|Vb_1\rangle|\gamma_{Vb_1}\rangle$, $|Vb_2\rangle|\gamma_{Vb_2}\rangle$, $|Hb_1\rangle|\theta_{Hb_1}\rangle$, $|Hb_2\rangle|\theta_{Hb_2}\rangle$, $|Vb_1\rangle|\theta_{Vb_1}\rangle$ and $|Vb_2\rangle|\theta_{Vb_2}\rangle$. According to Eqs.(5-8) and Eqs.(19-22), $\hat{U}_F$ automatically keeps the state of single photon after Bob's measurement unchanged. Further, according to Eqs.(9-10), Eqs.(23-24) and Eq.(44), Alice only can randomly obtain the measurement results $|Hb_1\rangle$ and $|Vb_1\rangle$ after her measurement with the $Z_P \otimes Z_S$ basis. Hence, in this situation, Eve is not detectable in Step 3.

It can be concluded that, in this circumstance, for Eve not inducing errors in Step 3, the final state of Eve's probe should be independent of Bob's choice of operation.

(3) Assume that $|\phi\rangle = |\phi\rangle_P \otimes |\phi\rangle_S$ is in the state of $|R\rangle \otimes |b_2\rangle$

The global state of the composite system formed by the single photon $|R\rangle \otimes |b_2\rangle$ and Eve's auxiliary particle $|\varepsilon\rangle$ before Eve's attack can be represented as $(|R\rangle \otimes |b_2\rangle) \otimes |\varepsilon\rangle$. According to the linearity of quantum mechanics together with Eq.(11) and Eq.(25), after Eve implements $\hat{U}_E$, the global state is turned into

$$\hat{U}_E((|R\rangle \otimes |b_2\rangle) \otimes |\varepsilon\rangle) = \hat{U}_E\left(\left(\frac{1}{\sqrt{2}}(|H\rangle + |V\rangle) \otimes |b_2\rangle\right) \otimes |\varepsilon\rangle\right)$$

$$= \frac{1}{\sqrt{2}}(|Hb_1\rangle|\mu_{Hb_1}\rangle + |Hb_2\rangle|\mu_{Hb_2}\rangle + |Vb_1\rangle|\mu_{Vb_1}\rangle + |Vb_2\rangle|\mu_{Vb_2}\rangle)$$

$$+ \frac{1}{\sqrt{2}}(|Hb_1\rangle|\sigma_{Hb_1}\rangle + |Hb_2\rangle|\sigma_{Hb_2}\rangle + |Vb_1\rangle|\sigma_{Vb_1}\rangle + |Vb_2\rangle|\sigma_{Vb_2}\rangle). \tag{47}$$

After Bob obtains the state from Alice, he randomly chooses either to CTRL or to SIFT. Eve imposes $\hat{U}_F$ on the state returned to Alice.

i) Consider the situation that Bob chooses to CRTL. Consequently, the global state is kept intact.

After Eve imposes $\hat{U}_F$ on the state returned to Alice, due to the linearity of quantum mechanics together with Eq.(17) and Eq.(31), the global state of Eq.(47) is turned into

$$\hat{U}_F\left(\hat{U}_E\left(\left(|R\rangle\otimes|b_2\rangle\right)\otimes|\varepsilon\rangle\right)\right) = \frac{1}{\sqrt{2}}\hat{U}_F\left(|Hb_1\rangle|\mu_{Hb_1}\rangle + |Hb_2\rangle|\mu_{Hb_2}\rangle + |Vb_1\rangle|\mu_{Vb_1}\rangle + |Vb_2\rangle|\mu_{Vb_2}\rangle\right)$$

$$+ \frac{1}{\sqrt{2}}\hat{U}_F\left(|Hb_1\rangle|\sigma_{Hb_1}\rangle + |Hb_2\rangle|\sigma_{Hb_2}\rangle + |Vb_1\rangle|\sigma_{Vb_1}\rangle + |Vb_2\rangle|\sigma_{Vb_2}\rangle\right)$$

$$= \frac{1}{\sqrt{2}}|Hb_2\rangle|\nu\rangle + \frac{1}{\sqrt{2}}|Vb_2\rangle|\tau\rangle. \tag{48}$$

For Eve not being discovered in Step 3, Alice's measurement result should be $|R\rangle\otimes|b_2\rangle$. Thus, it can be obtained from Eq.(48) that

$$|\nu\rangle = |\tau\rangle. \tag{49}$$

ii) Consider the situation that Bob chooses to SIFT. Consequently, the global state is collapsed into either of $|Hb_1\rangle|\mu_{Hb_1}\rangle$, $|Hb_2\rangle|\mu_{Hb_2}\rangle$, $|Vb_1\rangle|\mu_{Vb_1}\rangle$, $|Vb_2\rangle|\mu_{Vb_2}\rangle$, $|Hb_1\rangle|\sigma_{Hb_1}\rangle$, $|Hb_2\rangle|\sigma_{Hb_2}\rangle$, $|Vb_1\rangle|\sigma_{Vb_1}\rangle$ and $|Vb_2\rangle|\sigma_{Vb_2}\rangle$. According to Eqs.(12-15) and Eqs.(26-29), $\hat{U}_F$ automatically keeps the state of single photon after Bob's measurement unchanged. Further, according to Eqs.(16-17), Eqs.(30-31) and Eq.(49), Alice only can randomly obtain the measurement results $|Hb_2\rangle$ and $|Vb_2\rangle$ after her measurement with the $Z_P \otimes Z_S$ basis. Hence, in this situation, Eve is not detectable in Step 3.

It can be concluded that, in this circumstance, for Eve not inducing errors in Step 3, the final state of Eve's probe should be independent of Bob's choice of operation.

(4) Assume that $|\phi\rangle = |\phi\rangle_P \otimes |\phi\rangle_S$ is in the state of $|A\rangle\otimes|b_2\rangle$

The global state of the composite system formed by the single photon $|A\rangle\otimes|b_2\rangle$ and Eve's auxiliary particle $|\varepsilon\rangle$ before Eve's attack can be represented as $\left(|A\rangle\otimes|b_2\rangle\right)\otimes|\varepsilon\rangle$. According to the linearity of quantum mechanics together with Eq.(11) and Eq.(25), after Eve implements $\hat{U}_E$, the global state is turned into

$$\hat{U}_E\left(\left(|A\rangle\otimes|b_2\rangle\right)\otimes|\varepsilon\rangle\right) = \hat{U}_E\left(\left(\frac{1}{\sqrt{2}}\left(|H\rangle - |V\rangle\right)\otimes|b_2\rangle\right)\otimes|\varepsilon\rangle\right)$$

$$= \frac{1}{\sqrt{2}}\left(|Hb_1\rangle|\mu_{Hb_1}\rangle + |Hb_2\rangle|\mu_{Hb_2}\rangle + |Vb_1\rangle|\mu_{Vb_1}\rangle + |Vb_2\rangle|\mu_{Vb_2}\rangle\right)$$

$$- \frac{1}{\sqrt{2}}\left(|Hb_1\rangle|\sigma_{Hb_1}\rangle + |Hb_2\rangle|\sigma_{Hb_2}\rangle + |Vb_1\rangle|\sigma_{Vb_1}\rangle + |Vb_2\rangle|\sigma_{Vb_2}\rangle\right). \tag{50}$$

After Bob obtains the state from Alice, he randomly chooses either to CTRL or to SIFT. Eve imposes $\hat{U}_F$ on the state returned to Alice.

i) Consider the situation that Bob chooses to CRTL. Consequently, the global state is kept intact.

After Eve imposes $\hat{U}_F$ on the state returned to Alice, due to the linearity of quantum mechanics together with Eq.(17) and Eq.(31), the global state of Eq.(50) is turned into

$$\hat{U}_F\left(\hat{U}_E\left(\left(|A\rangle\otimes|b_2\rangle\right)\otimes|\varepsilon\rangle\right)\right) = \frac{1}{\sqrt{2}}\hat{U}_F\left(|Hb_1\rangle|\mu_{Hb_1}\rangle + |Hb_2\rangle|\mu_{Hb_2}\rangle + |Vb_1\rangle|\mu_{Vb_1}\rangle + |Vb_2\rangle|\mu_{Vb_2}\rangle\right)$$

$$- \frac{1}{\sqrt{2}}\hat{U}_F\left(|Hb_1\rangle|\sigma_{Hb_1}\rangle + |Hb_2\rangle|\sigma_{Hb_2}\rangle + |Vb_1\rangle|\sigma_{Vb_1}\rangle + |Vb_2\rangle|\sigma_{Vb_2}\rangle\right)$$

$$= \frac{1}{\sqrt{2}}|Hb_2\rangle|\nu\rangle - \frac{1}{\sqrt{2}}|Vb_2\rangle|\tau\rangle. \tag{51}$$

For Eve not being discovered in Step 3, Alice's measurement result should be $|A\rangle\otimes|b_2\rangle$. It is naturally derived after Eq.(49) is inserted into Eq.(51).

ii) Consider the situation that Bob chooses to SIFT. Consequently, the global state is collapsed into either of $|Hb_1\rangle|\mu_{Hb_1}\rangle$,

$|Hb_2\rangle|\mu_{Hb_2}\rangle$, $|Vb_1\rangle|\mu_{Vb_1}\rangle$, $|Vb_2\rangle|\mu_{Vb_2}\rangle$, $|Hb_1\rangle|\sigma_{Hb_1}\rangle$, $|Hb_2\rangle|\sigma_{Hb_2}\rangle$, $|Vb_1\rangle|\sigma_{Vb_1}\rangle$ and $|Vb_2\rangle|\sigma_{Vb_2}\rangle$. According to Eqs.(12-15) and Eqs.(26-29), $\hat{U}_F$ automatically keeps the state of single photon after Bob's measurement unchanged. Further, according to Eqs.(16-17), Eqs.(30-31) and Eq.(49), Alice only can randomly obtain the measurement results $|Hb_2\rangle$ and $|Vb_2\rangle$ after her measurement with the $Z_P \otimes Z_S$ basis. Hence, in this situation, Eve is not detectable in Step 3.

It can be concluded that, in this circumstance, for Eve not inducing errors in Step 3, the final state of Eve's probe should be independent of Bob's choice of operation.

**Case 4:** $|\phi\rangle = |\phi\rangle_P \otimes |\phi\rangle_S$ **is within the** $X_P \otimes X_S$ **basis**

(1) Assume that $|\phi\rangle = |\phi\rangle_P \otimes |\phi\rangle_S$ is in the state of $|R\rangle \otimes |s\rangle$

The global state of the composite system formed by the single photon $|R\rangle \otimes |s\rangle$ and Eve's auxiliary particle $|\varepsilon\rangle$ before Eve's attack can be represented as $(|R\rangle \otimes |s\rangle) \otimes |\varepsilon\rangle$. According to the linearity of quantum mechanics together with Eq.(4), Eq.(11), Eq.(18) and Eq.(25), after Eve implements $\hat{U}_E$, the global state is turned into

$$\hat{U}_E((|R\rangle \otimes |s\rangle) \otimes |\varepsilon\rangle) = \hat{U}_E\left(\left(\frac{1}{\sqrt{2}}(|H\rangle + |V\rangle) \otimes \frac{1}{\sqrt{2}}(|b_1\rangle + |b_2\rangle)\right) \otimes |\varepsilon\rangle\right)$$

$$= \frac{1}{2}(|Hb_1\rangle|\gamma_{Hb_1}\rangle + |Hb_2\rangle|\gamma_{Hb_2}\rangle + |Vb_1\rangle|\gamma_{Vb_1}\rangle + |Vb_2\rangle|\gamma_{Vb_2}\rangle)$$

$$+ \frac{1}{2}(|Hb_1\rangle|\mu_{Hb_1}\rangle + |Hb_2\rangle|\mu_{Hb_2}\rangle + |Vb_1\rangle|\mu_{Vb_1}\rangle + |Vb_2\rangle|\mu_{Vb_2}\rangle)$$

$$+ \frac{1}{2}(|Hb_1\rangle|\theta_{Hb_1}\rangle + |Hb_2\rangle|\theta_{Hb_2}\rangle + |Vb_1\rangle|\theta_{Vb_1}\rangle + |Vb_2\rangle|\theta_{Vb_2}\rangle)$$

$$+ \frac{1}{2}(|Hb_1\rangle|\sigma_{Hb_1}\rangle + |Hb_2\rangle|\sigma_{Hb_2}\rangle + |Vb_1\rangle|\sigma_{Vb_1}\rangle + |Vb_2\rangle|\sigma_{Vb_2}\rangle). \quad (52)$$

After Bob obtains the state from Alice, he randomly chooses either to CTRL or to SIFT. Eve imposes $\hat{U}_F$ on the state returned to Alice.

i) Consider the situation that Bob chooses to CRTL. Consequently, the global state is kept intact.

After Eve imposes $\hat{U}_F$ on the state returned to Alice, due to the linearity of quantum mechanics together with Eq.(10), Eq.(17), Eq.(24) and Eq.(31), the global state of Eq.(52) is turned into

$$\hat{U}_F\left(\hat{U}_E((|R\rangle \otimes |s\rangle) \otimes |\varepsilon\rangle)\right) = \frac{1}{2}\hat{U}_F(|Hb_1\rangle|\gamma_{Hb_1}\rangle + |Hb_2\rangle|\gamma_{Hb_2}\rangle + |Vb_1\rangle|\gamma_{Vb_1}\rangle + |Vb_2\rangle|\gamma_{Vb_2}\rangle)$$

$$+ \frac{1}{2}\hat{U}_F(|Hb_1\rangle|\mu_{Hb_1}\rangle + |Hb_2\rangle|\mu_{Hb_2}\rangle + |Vb_1\rangle|\mu_{Vb_1}\rangle + |Vb_2\rangle|\mu_{Vb_2}\rangle)$$

$$+ \frac{1}{2}\hat{U}_F(|Hb_1\rangle|\theta_{Hb_1}\rangle + |Hb_2\rangle|\theta_{Hb_2}\rangle + |Vb_1\rangle|\theta_{Vb_1}\rangle + |Vb_2\rangle|\theta_{Vb_2}\rangle)$$

$$+ \frac{1}{2}\hat{U}_F(|Hb_1\rangle|\sigma_{Hb_1}\rangle + |Hb_2\rangle|\sigma_{Hb_2}\rangle + |Vb_1\rangle|\sigma_{Vb_1}\rangle + |Vb_2\rangle|\sigma_{Vb_2}\rangle)$$

$$= \frac{1}{2}|Hb_1\rangle|\lambda\rangle + \frac{1}{2}|Hb_2\rangle|\nu\rangle + \frac{1}{2}|Vb_1\rangle|\vartheta\rangle + \frac{1}{2}|Vb_2\rangle|\tau\rangle. \quad (53)$$

Combining Eq.(34), Eq.(39), Eq.(44) and Eq.(49), we have

$$|\lambda\rangle = |\nu\rangle = |\vartheta\rangle = |\tau\rangle = |\omega\rangle. \quad (54)$$

After inserting Eq.(54) into Eq.(53), we have

$$\hat{U}_F\left(\hat{U}_E((|R\rangle \otimes |s\rangle) \otimes |\varepsilon\rangle)\right) = (|R\rangle \otimes |s\rangle) \otimes |\omega\rangle, \quad (55)$$

which can guarantee Eve not being detectable in Step 3, since Alice's measurement result is $|R\rangle \otimes |s\rangle$.

ii) Consider the situation that Bob chooses to SIFT. Consequently, the global state is collapsed into either of $|Hb_1\rangle|\gamma_{Hb_1}\rangle$,

$\left|Hb_2\right\rangle\left|\gamma_{Hb_2}\right\rangle$, $\left|Vb_1\right\rangle\left|\gamma_{Vb_1}\right\rangle$, $\left|Vb_2\right\rangle\left|\gamma_{Vb_2}\right\rangle$, $\left|Hb_1\right\rangle\left|\mu_{Hb_1}\right\rangle$, $\left|Hb_2\right\rangle\left|\mu_{Hb_2}\right\rangle$, $\left|Vb_1\right\rangle\left|\mu_{Vb_1}\right\rangle$, $\left|Vb_2\right\rangle\left|\mu_{Vb_2}\right\rangle$, $\left|Hb_1\right\rangle\left|\theta_{Hb_1}\right\rangle$, $\left|Hb_2\right\rangle\left|\theta_{Hb_2}\right\rangle$, $\left|Vb_1\right\rangle\left|\theta_{Vb_1}\right\rangle$, $\left|Vb_2\right\rangle\left|\theta_{Vb_2}\right\rangle$, $\left|Hb_1\right\rangle\left|\sigma_{Hb_1}\right\rangle$, $\left|Hb_2\right\rangle\left|\sigma_{Hb_2}\right\rangle$, $\left|Vb_1\right\rangle\left|\sigma_{Vb_1}\right\rangle$ and $\left|Vb_2\right\rangle\left|\sigma_{Vb_2}\right\rangle$. According to Eqs.(5-8), Eqs.(12-15), Eqs.(19-22) and Eqs.(26-29), $\hat{U}_F$ automatically keeps the state of single photon after Bob's measurement unchanged. Further, according to Eqs.(9-10), Eqs.(16-17), Eqs.(23-24), Eqs.(30-31) and Eq.(54), Alice only can randomly obtain the measurement results $\left|Hb_1\right\rangle$, $\left|Hb_2\right\rangle$, $\left|Vb_1\right\rangle$ and $\left|Vb_2\right\rangle$ after her measurement with the $Z_P \otimes Z_S$ basis. Hence, in this situation, Eve is not detectable in Step 3.

It can be concluded that, in this circumstance, for Eve not inducing errors in Step 3, the final state of Eve's probe should be independent of Bob's choice of operation.

(2) Assume that $\left|\phi\right\rangle = \left|\phi\right\rangle_P \otimes \left|\phi\right\rangle_S$ is in the state of $\left|R\right\rangle \otimes \left|a\right\rangle$, $\left|A\right\rangle \otimes \left|s\right\rangle$ or $\left|A\right\rangle \otimes \left|a\right\rangle$

When the single photon from Alice to Bob is in the state of $\left|R\right\rangle \otimes \left|a\right\rangle$, $\left|A\right\rangle \otimes \left|s\right\rangle$ or $\left|A\right\rangle \otimes \left|a\right\rangle$, after the similar deduction process as above, we can also draw the conclusion that for Eve not inducing errors in Step 3, the final state of Eve's probe should be independent of Bob's choice of operation.

After summarizing Cases 1, 2, 3 and 4, it can be easily derived that, for no error is caused by this attack in Step 3, the final state of Eve's probe should be irrelevant to Bob's choice of operation and the state in Bob's hand. As a result, Eve gets nothing useful about the final shared key bits. Therefore, the transmissions of single photons between Alice and Bob are completely robust. In other words, the proposed SQKD protocol is complete robust.

Thirdly, we consider how to defeat the Trojan horse attacks from Eve for this protocol, containing the invisible photon eavesdropping attack [19] and the delay-photon Trojan horse attack [20-21]. In accordance with Refs.[21-22], Bob can prevent the former attack by employing a filter in front of his devices and the latter attack by utilizing a photon number splitter (PNS).

## 5 Discussions and conclusions

Now we compare the proposed protocol with Boyer *et al.*'s famous pioneering SQKD protocol in Ref.[2] and the only existing SQKD protocol with single photons in two degrees of freedom in Ref.[18]. The comparison results are summarized in Table 1, after the classical bits needed for the security check processes are ignored. Here, the quantum communication efficiency is characterized as [23] $\eta = \frac{b_k}{b_q + b_c} \times 100\%$, where $b_k$, $b_q$ and $b_c$ are the expected number of private key bits established, the number of consumed qubits and the number of classical bits needed, respectively.

Table 1 Comparison results among the proposed protocol, the protocol of Ref.[2] and the protocol of Ref.[18]

| | $b_k$ | $b_q$ | $b_c$ | $\eta$ | $c_q$ | Number of kinds of initial quantum states | Whether the classical communicant need a quantum memory or a unitary operation equipment | Whether suffering from the double CNOT attack from Eve |
|---|---|---|---|---|---|---|---|---|
| The protocol of Ref.[2] | $n$ | $12n(1+\delta)$ | 0 | 8.33% | 1 | Four | No | No |
| The protocol of Ref.[18] | $n$ | $12n(1+\delta)$ | 0 | 8.33% | 2 | One | No | No |
| The proposed protocol | $n$ | $4.5n(1+\delta)$ | 0 | 11.11% | 2 | Sixteen | No | No |

In the protocol of Ref.[2], for establishing $n$ private key bits between quantum Alice and classical Bob, Alice needs to generate $8n(1+\delta)$ polarized single photons in one degree of freedom, while Bob needs to prepare $4n(1+\delta)$ ones when he chooses to SIFT. There are no classical bits needed for helping establish the private key bits. Hence, we have $b_k = n$, $b_q = 8n(1+\delta) + 4n(1+\delta) = 12n(1+\delta)$ and $b_c = 0$. As a result, the efficiency of the protocol of Ref.[2] is $\eta = \frac{n}{12n(1+\delta)} \times 100\% \approx 8.33\%$, since $\delta$ is always small enough to be neglected.

In the protocol of Ref.[18], for establishing $n$ private key bits between quantum Alice and classical Bob, Alice needs to generate $4n(1+\delta)$ single photons in two degrees of freedom, while Bob needs to prepare $2n(1+\delta)$ ones when he chooses to SIFT. There are no classical bits needed for helping establish the private key bits. Hence, we have $b_k = n$, $b_q = 4n(1+\delta) \times 2 + 2n(1+\delta) \times 2 = 12n(1+\delta)$ and $b_c = 0$. As a result, the efficiency of the protocol of Ref.[18] is $\eta = \frac{n}{12n(1+\delta)} \times 100\% \approx 8.33\%$.

In the proposed protocol, for establishing $n$ private key bits between quantum Alice and classical Bob, Alice needs to generate $1.5n(1+\delta)$, $0.5n(1+\delta)$, $0.5n(1+\delta)$ and $0.5n(1+\delta)$ single photons in two degrees of freedom randomly in the $Z_P \otimes Z_S$ basis, the $Z_P \otimes X_S$ basis, the $X_P \otimes Z_S$ basis and the $X_P \otimes X_S$ basis, respectively, while Bob needs to prepare $1.5n(1+\delta)$ ones when he chooses to SIFT. There are no classical bits needed for helping establish the private key bits. Hence, we have $b_k = n$, $b_q = 1.5n(1+\delta) + 0.5n(1+\delta) \times 3 + 1.5n(1+\delta) = 4.5n(1+\delta)$ and $b_c = 0$. As a result, the efficiency of the proposed protocol is $\eta = \frac{n}{4.5n(1+\delta) \times 2} \times 100\% \approx 11.11\%$.

It can be concluded from Table 1 that:

①In the protocol of Ref.[2], one single photon in one degree of freedom for establishing the private key bits always carries one private bit, while in the proposed protocol, one single photon in two degrees of freedom for establishing the private key bits always carries two private bits. Therefore, the quantum communication capacity $c_q$ of the proposed protocol is twice that of the protocol in Ref.[2];

②The proposed protocol exceeds the protocol of Ref.[2] and the protocol of Ref.[18] in quantum communication efficiency, as it consumes less qubits for establishing a private key of the same length.

To sum up, in this paper, an efficient SQKD protocol with single photons in both polarization and spatial-mode degrees of freedom is suggested, which is feasible for a quantum communicant distributing a random private key to a classical communicant. The proposed protocol needn't require the classical communicant to employ any quantum memory or unitary operation equipment. The complete robustness of the transmissions of single photons between two communicants is validated. Compared with Boyer *et al.*'s famous pioneering SQKD protocol in Ref.[2], this protocol has double quantum communication capacity and higher quantum communication efficiency. Compared with the only existing SQKD protocol with single photons in two degrees of freedom in Ref.[18], this protocol has higher quantum communication efficiency. In the future, we will study how to design other semiquantum cryptography protocols based on single photons in two degrees of freedom.

**Acknowledgments**

The authors would like to thank the anonymous reviewers for their valuable comments that help enhancing the quality of this paper. Funding by the National Natural Science Foundation of China (Grant No.62071430 and No.61871347), the Fundamental Research Funds for the Provincial Universities of Zhejiang (Grant No.JRK21002) and Zhejiang Gongshang University, Zhejiang Provincial Key Laboratory of New Network Standards and Technologies (No. 2013E10012) is gratefully acknowledged.

**Data Availability Statement**

All data and models generated or used during the study appear in the submitted article.